\documentclass[aps,prb,amsmath,twocolumn,showpacs]{revtex4}
\usepackage{bm}
\usepackage{graphicx}

\DeclareMathOperator{\Img}{\mathrm{Im}}
\DeclareMathOperator{\Rea}{\mathrm{Re}}
\DeclareMathOperator{\Sp}{\mathrm{Sp}}

\begin{document}
\title{Non-local spin-sensitive electron transport in diffusive proximity heterostructures}
\author{Mikhail S. Kalenkov}
\affiliation{I.E. Tamm Department of Theoretical Physics, P.N.
Lebedev Physics Institute, 119991 Moscow, Russia}
\author{Andrei D. Zaikin}
\affiliation{Institute for Nanotechnology, Karlsruhe Institute of
Technology (KIT), 76021 Karlsruhe, Germany} \affiliation{I.E. Tamm
Department of Theoretical Physics, P.N. Lebedev Physics Institute,
119991 Moscow, Russia}

\begin{abstract}
We formulate a quantitative theory of non-local electron transport in three-terminal disordered ferromagnet-superconductor-ferromagnet structures. We demonstrate that magnetic effects have different
implications: While strong exchange field suppresses disorder-induced electron interference in ferromagnetic electrodes, spin-sensitive electron scattering at superconductor-ferromagnet interfaces can drive the total non-local conductance $g_{12}$ negative at sufficiently low energies.
At higher energies magnetic effects become less important and the non-local resistance behaves similarly to the non-magnetic case.
Our predictions can be directly tested in
future experiments on non-local electron transport in hybrid $FSF$ structures.
\end{abstract}



\maketitle
\section{Introduction}

The phenomenon of Andreev reflection (AR) \cite{And} is well known
to be responsible for transport of subgap electrons across an
interface between a normal metal ($N$) and a superconductor ($S$).
While this phenomenon is essentially local in hybrid proximity
structures with only one $NS$ interface, the situation in
multiterminal devices with two or more $NS$ interfaces (such as,
e.g., $NSN$ structures) can be more complicated because in
addition to local AR electrons can suffer non-local or crossed
Andreev reflection (CAR) \cite{car}. This phenomenon of CAR enables
direct experimental demonstration of entanglement between
electrons in spatially separated $N$-electrodes and can strongly
influence non-local transport of electrons in hybrid $NSN$ systems
\cite{FFH,KZ06}.

Non-local electron transport in the presence of CAR was recently
investigated both experimentally
\cite{Beckmann,Teun,Venkat,Basel,Deutscher,Beckmann2} and theoretically
\cite{FFH,KZ06,LY,GZ09,BG,Belzig,Melin,GZ07,Golubev09,Kalenkov07,Kalenkov07E}
demonstrating a rich variety of physical processes involved in the
problem. For instance, the effect of CAR on the subgap non-local
conductance of $NSN$ structures is exactly compensated by elastic
cotunneling (EC) provided only the lowest order terms in $NS$
interface transmissions  are accounted for \cite{FFH}.  Taking
into account higher order processes in barrier transmissions
eliminates this feature and yields non-zero values of
cross-conductance \cite{KZ06}. One can also expect that
interactions \cite{LY} or external ac bias \cite{GZ09} can lift
the cancellation between EC and CAR contributions already in the
lowest order in barrier transmissions.

Another non-trivial issue is the effect of disorder. Theoretical
analysis of CAR in different disordered $NSN$ structures was
carried out in Refs. \onlinecite{BG,Belzig,Melin,GZ07,Golubev09}. In
particular, it was demonstrated \cite{Golubev09} that an interplay
between CAR, quantum interference of electrons and non-local
charge imbalance dominates the behavior of diffusive $NSN$ systems
being essential for quantitative interpretation of a number of
experimental observations \cite{Venkat,Basel,Deutscher}.

Yet another important property of both local and non-local Andreev
reflection processes is that they essentially depend on spins of
scattered electrons. Hence, CAR should be sensitive to magnetic
properties of normal electrodes. This sensitivity was indeed
demonstrated already in the first experiments on
ferromagnet-superconductor-ferromagnet ($FSF$) structures
\cite{Beckmann} where the dependence of non-local conductance on
the polarization of ferromagnetic terminals was found. Theoretical
analysis of spin-resolved CAR was carried out in Ref.
\onlinecite{FFH} in the lowest order order in tunneling and in
Refs. \onlinecite{Kalenkov07}, \onlinecite{Kalenkov07E} to all orders in the interface
transmissions. This analysis revealed a number of non-trivial
features of non-local spin-dependent electron transport which can
be tested in future experiments.

Note that previous work \cite{FFH,Kalenkov07,Kalenkov07E} merely concentrated
on ballistic electrodes whereas in realistic experiments one
usually deals with diffusive hybrid $FSF$ structures. Therefore it
is highly desirable to formulate a theory which would adequately
describe an interplay between disorder and spin-resolved CAR. This
is the main goal of the present paper. The structure of our paper
is as follows. In Sec. 2 we will formulate our model and outline
our basic formalism of quasiclassical Green functions. This
formalism will be employed in Sec. 3 where we present the solution
of Usadel equations and derive general expressions for
the non-local spin-dependent conductance and resistance for diffusive
three-terminal $FSF$ structures at different directions of interface magnetizations.  Concluding remarks are presented in Sec. 4
of our paper.

\section{Model and basic formalism}
\begin{figure}
\includegraphics[width=7.5cm]{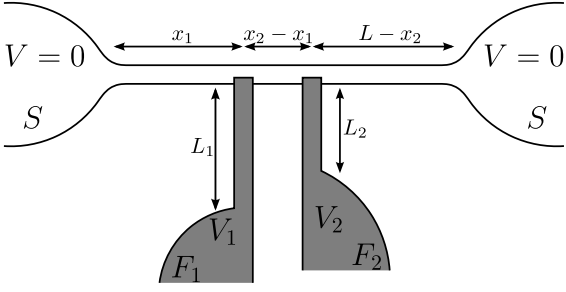}
\caption{FSF structure under consideration.}
\label{ufsf-fig}
\end{figure}

Let us consider a three-terminal diffusive $FSF$ structure
schematically shown in Fig. \ref{ufsf-fig}. Two ferromagnetic terminals $F_1$
and $F_2$ with resistances $r_{N_1}$ and $r_{N_2}$ and electric
potentials $V_1$ and $V_2$ are connected to a superconducting
electrode of length $L$ with normal state (Drude) resistance $r_L$
and electric potential $V=0$ via tunnel barriers. The magnitude of
the exchange field $h_{1,2}=|\bm{h}_{1,2}|$ in both ferromagnets $F_1$ and
$F_2$ is assumed to be much bigger than the superconducting order
parameter $\Delta$ of the $S$-terminal and, on the other hand,
much smaller that the Fermi energy, i.e. $\Delta \ll h_{1,2} \ll
\epsilon_F$.

The latter condition allows to perform the analysis of our $FSF$
system within the quasiclassical formalism of Usadel equations for
the Green-Keldysh matrix functions $G$. In each of our metallic
terminals these equations can be written in the form \cite{BWBSZ}
\begin{equation}
iD\nabla (\check G \nabla \check G)=
[\check\Omega +eV , \check G], \quad \check G^2=1,
\label{Usadel}
\end{equation}
where $D$ is the diffusion constant, $V$ is the electric potential, $\check G$ and $\check\Omega$ are $8\times8$ matrices in Keldysh-Nambu-spin space (denoted by check symbol)
\begin{gather}
\check G=
\begin{pmatrix}
\breve G^R & \breve G^K \\
0 & \breve G^A \\
\end{pmatrix}, \quad
\check \Omega=
\begin{pmatrix}
\breve \Omega^R & 0 \\
0 & \breve \Omega^A \\
\end{pmatrix},
\\
\breve \Omega^R=\breve \Omega^A=
\begin{pmatrix}
\varepsilon - \hat{\bm{\sigma}}\bm{h} & \Delta \\
-\Delta^* & -\varepsilon + \hat{\bm{\sigma}}\bm{h}\\
\end{pmatrix},
\end{gather}
$\varepsilon$ is the quasiparticle energy, $\Delta (T)$ is
the superconducting order parameter which will be considered real
in a superconductor and zero in both ferromagnets, $\bm{h}\equiv\bm{h}_{1(2)}$ in the
first (second) ferromagnetic terminal, $\bm{h}\equiv 0$ outside these terminals and $\hat{\bm{\sigma}}=(\hat\sigma_1,\hat\sigma_2,\hat\sigma_3)$ are Pauli matrices in spin space.

Retarded and advanced Green functions $\breve G^R$ and $\breve G^A$ have the following matrix structure
\begin{equation}
\breve G^{R,A}=
\begin{pmatrix}
\hat G^{R,A}  & \hat F^{R,A} \\
-\hat F^{R,A} & -\hat G^{R,A} \\
\end{pmatrix}.
\end{equation}
Here and below $2\times 2$ matrices in spin space are denoted by hat symbol.

Having obtained the expressions for the Green-Keldysh functions $\check G$ one can easily evaluate the current density $\bm{j}$ in our system
with the aid of the standard relation
\begin{equation}
\bm{j}= -\frac{\sigma}{16e} \int \Sp [ \tau_3 ( \check G \nabla  \check G)^K]d
\varepsilon ,\label{current}
\end{equation}
where $\sigma$ is the Drude conductivity of the corresponding
metal and $\tau_3$ is the Pauli matrix in Nambu space.

In what follows it will be convenient for us to employ the
so-called Larkin-Ovchinnikov parameterization of the Keldysh Green
function
\begin{equation}
\breve G^K=\breve G^R \breve f - \breve f \breve G^A,
\quad \breve f = \hat f_L + \tau_3 \hat f_T,
\end{equation}
where the distribution functions $\hat f_L$ and $\hat f_T$ are $2\times2$
matrices in the spin space.

For the sake of simplicity we will assume that
magnetizations of both ferromagnets and the interfaces (see below)
are collinear. Within this approximation the Green functions and
the matrix $\check \Omega$ are diagonal in the spin space and the
diffusion-like equations for the distribution function matrices
$\hat f_L$ and $\hat f_T$ take the form
\begin{gather}
- D \nabla \left( \hat D^T(\bm{r},\varepsilon) \nabla
\hat f_T(\bm{r},\varepsilon) \right) + 2\hat \Sigma(\bm{r},\varepsilon)
\hat f_T(\bm{r},\varepsilon) =0, \label{diffeqT}
\\
- D \nabla \left( \hat D^L(\bm{r},\varepsilon) \nabla
\hat f_L(\bm{r},\varepsilon) \right) =0, \label{diffeqL}
\end{gather}
where
\begin{gather}
\hat \Sigma (\bm{r},\varepsilon)= - i \Delta \Img \hat F^R,
\\
\hat D^T =\left( \Rea \hat G^R\right)^2 + \left( \Img \hat F^R\right)^2,
\\
\hat D^L =\left( \Rea \hat G^R\right)^2 - \left( \Rea \hat F^R\right)^2.
\end{gather}
The function $\hat \Sigma (\bm{r},\varepsilon)$ differs from zero only
inside the superconductor. It accounts both for energy relaxation of
quasiparticles and for their conversion to Cooper pairs due to Andreev
reflection. The functions $\hat D^T$ and $\hat D^L$ acquire space and energy
dependencies due to the presence of the superconducting wire and
renormalize the diffusion coefficient $D$.

The solution of Eqs. \eqref{diffeqT}-\eqref{diffeqL} can be
expressed in terms of the diffuson-like functions $\hat{\mathcal{D}}^T$
and $\hat{\mathcal{D}}^L$ which obey the following equations
\begin{gather}
\begin{split}
- D \nabla \left[
\hat D^T(\bm{r},\varepsilon) \nabla \hat{\mathcal{D}}^T(\bm{r},\bm{r}^{\prime},\varepsilon)
\right]  &
\\
+2\hat \Sigma (\bm{r},\varepsilon)
\hat{\mathcal{D}}^T(\bm{r},\bm{r}^{\prime},\varepsilon) =&
\delta(\bm{r}-\bm{r}^{\prime}),
\end{split}
\\
- D \nabla \left[
\hat D^L(\bm{r},\varepsilon) \nabla \hat{\mathcal{D}}^L(\bm{r},\bm{r}^{\prime},\varepsilon)
\right] =
\delta(\bm{r}-\bm{r}^{\prime}).
\end{gather}

The solutions of Usadel equation (\ref{Usadel}) in each of the metals
should be matched at $SF$-interfaces by means of appropriate
boundary conditions which account for electron tunneling between
these terminals. The form of these boundary conditions essentially
depends on the adopted model describing electron scattering at
$SF$-interfaces. Here we stick to the model of the so-called
spin-active interfaces \cite{Eschrig} which takes into account
possibly different barrier transmissions for spin-up and spin-down
electrons. This model was already extensively used for theoretical
description of different physical phenomena, including
spin-resolved CAR in ballistic structures \cite{Kalenkov07,Kalenkov07E} and
Josephson effect with triplet pairing \cite{Eschrig2,GKZ}. Here we
employ this model in the case of diffusive electrodes and also
restrict our analysis to the case of tunnel barriers with channel
transmissions much smaller than one. In this case the
corresponding boundary conditions read \cite{Huertas02}
\begin{gather}
\begin{split}
\mathcal{A}\sigma_+ & \check G_+  \partial_x \check G_+=
\dfrac{G_T}{2}
[\check G_-, \check G_+]
\\
& +\dfrac{G_m}{4}
[\{\hat{\bm{\sigma}}\bm{m} \tau_3, \check G_-\}, \check G_+] +
i\dfrac{G_{\varphi}}{2}
[\hat{\bm{\sigma}}\bm{m} \tau_3, \check G_+],
\label{gplus}
\end{split}
\\
\begin{split}
-\mathcal{A} & \sigma_-  \check G_-  \partial_x \check G_-=
\dfrac{G_T}{2}
[\check G_+, \check G_-]
\\
& +\dfrac{G_m}{4}
[\{\hat{\bm{\sigma}}\bm{m} \tau_3, \check G_+\}, \check G_-] +
i\dfrac{G_{\varphi}}{2}
[\hat{\bm{\sigma}}\bm{m} \tau_3, \check G_-],
\label{gminus}
\end{split}
\end{gather}
where $\check G_-$ and $\check G_+$ are the Green-Keldysh functions from the left ($x<0$) and from the right ($x>0$) side of the interface, $\mathcal{A}$ is the effective contact
area, $\bm{m}$ is the unit vector in the direction of the
interface magnetization, $\sigma_{\pm}$ are Drude
conductivities of the left and right terminals and $G_T$ is the
spin-independent part of the interface conductance. Along with
$G_T$ there also exists the spin-sensitive contribution to the
interface conductance which is accounted for by the $G_m$-term,
whereas the $G_{\varphi}$-term arises due to different phase
shifts acquired by scattered quasiparticles with opposite spin
directions.

Employing the above boundary conditions we can establish the
following linear relations between the distribution functions at
both sides of the interface
\begin{gather}
\begin{split}
\mathcal{A}\sigma_+ \hat D_+^T &\partial_x \hat f_{+T} =\mathcal{A}\sigma_-
\hat D_-^T \partial_x \hat f_{-T}
\\
&=\hat g_T (\hat f_{+T} - \hat f_{-T}) + \hat g_m (\hat f_{+L} - \hat f_{-L}) ,\label{boundT}
\end{split}
\\
\begin{split}
\mathcal{A}\sigma_+ \hat D_+^L &\partial_x \hat f_{+L}=\mathcal{A}\sigma_-
\hat D_-^L \partial_x \hat f_{-L}
\\
&=\hat g_L (\hat f_{+L} - \hat f_{-L}) + \hat g_m (\hat f_{+T} - \hat f_{-T}) ,\label{boundL}
\end{split}
\end{gather}
where $\hat g_T$, $\hat g_L$, and $\hat g_m$ are matrix interface conductances which
depend on the retarded and advanced Green functions at the
interface
\begin{gather}
\hat g_T= G_T \left[\left(\Rea \hat G^R_+\right) \left(\Rea \hat G^R_-\right)
+ \left(\Img \hat F^R_+\right)\left(\Img \hat F^R_-\right)\right],
\\
\hat g_L= G_T \left[ \left(\Rea \hat G^R_+\right) \left(\Rea \hat G^R_-\right) -
\left(\Rea \hat F^R_+\right)\left(\Rea \hat F^R_-\right)\right],
\\
\hat g_m = G_m \hat{\bm{\sigma}}\bm{m}\left(\Rea \hat G^R_+\right) \left(\Rea
\hat G^R_-\right).
\end{gather}

The current density \eqref{current} can then be expressed in terms
of the distribution function $\hat f_T$ as
\begin{equation}
\bm{j}= -\frac{\sigma}{4e} \int \Sp [ \hat D^T  \nabla  \hat f_T ]d
\varepsilon . \label{current1}
\end{equation}

\section{Spectral conductances}
Let us now employ the above formalism in order to evaluate electric currents in our
$FSF$ device depicted in Fig. 1. The current across the first ($SF_1$) interface can be written as
\begin{multline}
I_1 = \dfrac{1}{e}\int g_{11}(\varepsilon)\left[f_0(\varepsilon
+ eV_1)-f_0(\varepsilon)\right] d \varepsilon
\\
-\dfrac{1}{e}\int g_{12}(\varepsilon)\left[f_0(\varepsilon +
eV_2)-f_0(\varepsilon)\right] d \varepsilon, \label{intcur1}
\end{multline}
where $f_0(\varepsilon)=\tanh(\varepsilon/2T)$, $g_{11}$ and
$g_{12}$ are local and nonlocal spectral electric conductances.
Expression for the current across the second interface can be
obtained from the above equation by interchanging the indices
$1\leftrightarrow2$. Solving Eqs. \eqref{diffeqT}-\eqref{diffeqL}
with boundary conditions \eqref{boundT}-\eqref{boundL} we express
both local and nonlocal conductances $\hat g_{ij}(\varepsilon)$ in terms of
the interface conductances and the function $\hat{\mathcal{D}}$. The
corresponding results read
\begin{gather}
\begin{split}
\hat g_{11}(\varepsilon)= \bigl(
\hat R_2^T \hat{\mathcal{M}}^{L}+
& \hat R_2^T \hat R_2^L \hat R_{1m} - \hat R_1^L \hat R_{2m}^2
\\+
& \hat R_{12}^{T} \hat R_{12}^{L} \hat R_{2m}-
\hat R_{1m} \hat R_{2m}^2
\bigr)\hat{\mathcal{K}},
\end{split}
\label{g11}
\\
\begin{split}
\hat g_{12}(\varepsilon) = \hat g_{21}(\varepsilon) =\bigl(
& \hat R_{12}^T \hat {\mathcal{M}}^{L}+
\hat R_2^T \hat R_{12}^L \hat R_{1m}
\\+
& \hat R_{12}^{L} \hat R_{1m} \hat R_{2m} +
\hat R_{12}^{T} \hat R_{1}^L \hat R_{2m}
\bigr)\hat{\mathcal{K}},
\end{split}
\label{g12}
\end{gather}
where we defined
\begin{gather}
\hat{\mathcal{M}}^{T,L}=\hat R_1^{T,L} \hat R_2^{T,L}-(\hat R_{12}^{T,L})^2,
\\
\begin{split}
\hat{\mathcal{K}}^{-1}=
\hat{\mathcal{M}}^{T} \hat{\mathcal{M}}^{L}+
&\hat R_{1m}^2 \hat R_{2m}^2 -
\hat R_2^T \hat R_2^L \hat R_{1m}^2
\\-
2 &\hat R_{12}^{T} \hat R_{12}^{L} \hat R_{1m} \hat R_{2m} -
\hat R_1^T \hat R_1^L \hat R_{2m}^2
\end{split}
\label{calK}
\end{gather}
and introduced the auxiliary resistance matrix
\begin{multline}
\hat R_1^T=
\hat g_{1T}(\varepsilon)[
\hat g_{1T}(\varepsilon)\hat g_{1L}(\varepsilon) - \hat g_{1m}^2(\varepsilon)]^{-1}
\\+
\dfrac{D_1 \hat{\mathcal{D}}_1^T(\bm{r}_1,\bm{r}_1,\varepsilon)}{\sigma_1} +
\dfrac{D_S \hat{\mathcal{D}}_S^T(\bm{r}_1,\bm{r}_1,\varepsilon)}{\sigma_S},
\label{RT}
\end{multline}
The resistance matrices $\hat R_2^T$, $\hat R_1^L$ and $\hat R_2^L$ can be obtained by interchanging the indices $1\leftrightarrow2$ and $T\leftrightarrow L$ in Eq. (\ref{RT}). The remaining resistance matrices $\hat R_{12}^{T,L}$ and $\hat R_{jm}$ are defined as
\begin{gather}
\hat R_{12}^{T,L}= \hat R_{21}^{T,L}=
\dfrac{D_S \hat{\mathcal{D}}_S^{T,L}(\bm{r}_1,\bm{r}_2,\varepsilon)}{\sigma_S},
\\
\hat R_{jm}=\hat g_{jm}(\varepsilon)[
\hat g_{jT}(\varepsilon)\hat g_{jL}(\varepsilon) - \hat g_{jm}^2(\varepsilon)]^{-1},
\label{Rm}
\end{gather}
where $j=1,2$. The spectral conductance $g_{ij}$ can be recovered from
the matrix $\hat g_{ij}$ simply by summing up over the spin states
\begin{equation}
g_{ij}(\varepsilon)=\dfrac{1}{2}\Sp\left[\hat g_{ij}(\varepsilon)\right].
\end{equation}

It is worth pointing out that Eqs. \eqref{g11}, \eqref{g12} defining respectively  local and nonlocal spectral conductances are presented with excess accuracy. This is because the boundary conditions \eqref{gplus}-\eqref{gminus} employed here remain applicable only in the tunneling limit and for weak spin dependent scattering $|G_m|, |G_{\varphi}| \ll G_T$. Hence, strictly speaking only the lowest order terms in $G_{m_{1,2}}$ and $G_{\varphi_{1,2}}$ need to be kept in our final results.

In order to proceed it is necessary to evaluate the interface conductances as well as the matrix functions $\hat{\mathcal{D}}^{T,L}_{1,2,S}$. Restricting ourselves to the second order in the interface transmissions we obtain
\begin{gather}
\hat g_{1T}(\varepsilon)=G_{T_1} \hat \nu_S(\bm{r}_1,\varepsilon)
+ G_{T_1}^2 \dfrac{\Delta^2\theta(\Delta^2-\varepsilon^2)}{\Delta^2-\varepsilon^2}
\hat U_1(\varepsilon),
\label{g1T}
\\
\hat g_{1L}(\varepsilon)=G_{T_1}\hat \nu_S(\bm{r}_1,\varepsilon)
-G_{T_1}^2 \dfrac{\Delta^2\theta(\varepsilon^2-\Delta^2)}{\varepsilon^2-\Delta^2}
\hat U_1(\varepsilon),
\label{g1L}
\\
\hat g_{1m}(\varepsilon)=G_{m_1} \hat \nu_S(\bm{r}_1,\varepsilon)\hat{\bm{\sigma}}\bm{m}_1,
\label{g1m}
\end{gather}
and analogous expressions for the interface conductances of the second interface. The matrix function
\begin{multline}
\hat U_1(\varepsilon)= \dfrac{D_1}{2\sigma_1}\Bigl\{ \Rea \left[
\mathcal{C}_1(\bm{r}_1,\bm{r}_1,2h_1^+) +
\mathcal{C}_1(\bm{r}_1,\bm{r}_1,2h_1^-) \right]
\\-
\hat{\bm{\sigma}}\bm{m}_1 \Rea \left[
\mathcal{C}_1(\bm{r}_1,\bm{r}_1,2h_1^+)-
\mathcal{C}_1(\bm{r}_1,\bm{r}_1,2h_1^-) \right]
\Bigr\}
\end{multline}
with $h_1^{\pm}=h_1 \pm \varepsilon$ defines the correction due to the proximity effect in the normal metal.

Taking into account the first order corrections in the interface transmissions one can derive the density of states inside the superconductor in the following form
\begin{multline}
\hat \nu_S(\bm{r},\varepsilon)=
\dfrac{|\varepsilon| \theta (\varepsilon^2 - \Delta^2)}{
\sqrt{|\varepsilon^2 - \Delta^2|}}
\\
+\dfrac{D_S}{\sigma_S}\dfrac{\Delta^2}{\Delta^2 - \varepsilon^2}
\sum_{i=1,2}
\Biggl[
G_{T_i} \Rea \mathcal{C}_S(\bm{r},\bm{r}_i, 2\omega^R)
\\
-\hat{\bm{\sigma}}\bm{m}_i
G_{\varphi_i}\Img \mathcal{C}_S(\bm{r},\bm{r}_i, 2\omega^R)\Biggr],
\label{dos}
\end{multline}
where
\begin{equation}
\omega^R=
\begin{cases}
\sqrt{\varepsilon^2-\Delta^2}, \quad   &   \varepsilon>\Delta,
\\
i\sqrt{\Delta^2 -\varepsilon^2}, \quad &  |\varepsilon| < \Delta,
\\
-\sqrt{\varepsilon^2-\Delta^2}, \quad  &  \varepsilon < \Delta,
\end{cases}
\end{equation}
and the Cooperon $\mathcal{C}_{j}(\bm{r}, \bm{r}^{\prime}, \varepsilon)$ represents the solution of the equation
\begin{equation}
\left(-D\nabla^2-i\varepsilon\right)\mathcal{C}(\bm{r}, \bm{r}^{\prime}, \varepsilon)=
\delta(\bm{r}- \bm{r}^{\prime})
\end{equation}
in the normal metal leads ($j=1,2$) and the superconductor ($j=S$).
In the quasi-one-dimensional geometry the corresponding solutions take the form
\begin{gather}
\mathcal{C}_j (x_j, x_j, \varepsilon) =
\dfrac{\tanh\left(k_j L_j\right)}{S_jD_jk_j},
\quad j=1,2,
\\
\mathcal{C}_S (x,x',\varepsilon)=
\dfrac{\sinh [k_S(L-x')]\sinh k_S x}{k_S S_S D_S\sinh (k_S L)}, \quad x'>x,
\end{gather}
where $S_{S,1,2}$ are the wire cross sections and $k_{1,2,S}=\sqrt{-i\varepsilon/D_{1,2,S}}$.

Substituting Eq. (\ref{dos}) into Eqs. (\ref{g1T}) and (\ref{g1L}) and comparing the terms $\propto G_{T_1}^2$ we observe that the tunneling correction to the density of states dominates over the terms proportional to $\hat U_1$ which contain an extra small factor $\sqrt{\Delta/h} \ll 1$. Hence, the latter terms in Eqs. (\ref{g1T}) and (\ref{g1L}) can be safely neglected. In addition, in Eq. \eqref{dos} we also neglect small tunneling corrections to the superconducting density of states at energies exceeding the superconducting gap $\Delta$. Within this approximation the density of states inside the superconducting wire becomes spin-independent $\hat \nu_S (\bm{r} , \varepsilon) = \hat \sigma_0 \nu_S( \bm{r} , \varepsilon)$. It can then be written as
\begin{multline}
\nu_S(\bm{r},\varepsilon)=
\dfrac{|\varepsilon|}{\sqrt{|\varepsilon^2 - \Delta^2|}}
\theta(\varepsilon^2-\Delta^2)
\\
+\dfrac{D_S}{\sigma_S}\dfrac{\Delta^2\theta(\Delta^2-\varepsilon^2)}{\Delta^2 - \varepsilon^2}
\sum\limits_{i=1,2}
G_{T_i} \Rea \mathcal{C}_S(\bm{r},\bm{r}_i, 2\omega^R).
\label{dos1}
\end{multline}
Accordingly, the interface conductances take the form
\begin{gather}
\hat g_{1T}(\varepsilon) = \hat g_{1L}(\varepsilon)=G_{T_1}\nu_S(\bm{r}_1,\varepsilon),
\label{g1TL1}
\\
\hat g_{1m}(\varepsilon)=G_{m_1} \nu_S(\bm{r}_1,\varepsilon)\hat{\bm{\sigma}}\bm{m}_1.
\label{g1m1}
\end{gather}

In the limit of strong exchange fields $h_{1,2} \gg \Delta$ and small interface transmissions considered here the  proximity effect in the ferromagnets remains weak and can be neglected. Hence, the functions $\hat{\mathcal{D}}^{T,L}_{1}(\bm{r}_1,\bm{r}_1,\varepsilon)$ and $\hat{\mathcal{D}}^{T,L}_{2}(\bm{r}_2,\bm{r}_2,\varepsilon)$ can be approximated by their normal state values
\begin{gather}
\hat{\mathcal{D}}^{T,L}_{1}(\bm{r}_1,\bm{r}_1,\varepsilon)=\sigma_1 r_{N_1}\hat 1/D_1,
\\
\hat{\mathcal{D}}^{T,L}_{2}(\bm{r}_2,\bm{r}_2,\varepsilon)=\sigma_2 r_{N_2}\hat 1/D_2,
\\
r_{N_j} = L_j/(\sigma_j S_j), \quad j=1,2,
\end{gather}
where $r_{N_1}$ and $r_{N_2}$ are the normal state resistances of ferromagnetic terminals.
In the the superconducting region an effective expansion parameter is
$G_{T_{1,2}}r_{\xi_S}(\varepsilon)$, where
$r_{\xi_S}(\varepsilon)=\xi_S(\varepsilon)/(\sigma_S S_S)$ is the Drude
resistance of the superconducting wire segment of length $\xi_S(\varepsilon)=\sqrt{D_S/2|\omega^R|}$. In the limit
\begin{equation}
G_{T_{1,2}}r_{\xi_S}(\varepsilon) \ll 1, 
\end{equation}
which is typically well satisfied for realistic system parameters, it suffices to evaluate the function $\hat{\mathcal{D}}^T_S(x,x',\varepsilon)$ for impenetrable interfaces. In this case we find
\begin{equation}
\hat{\mathcal{D}}_S^T (x,x',\varepsilon)=
\begin{cases}
\dfrac{\Delta^2 - \varepsilon^2}{\Delta^2}\mathcal{C}_S (x,x',2\omega^R),  &|\varepsilon|  < \Delta,
\\
\dfrac{\varepsilon^2 -\Delta^2 }{\varepsilon^2}\mathcal{C}_S (x,x',0),  &|\varepsilon| > \Delta.
\end{cases}
\end{equation}
We note that special care should be taken while calculating $\mathcal{D}^L_S(x,x',\varepsilon)$ at subgap energies, since the coefficient $D^L$ in Eq. \eqref{diffeqL} tends to zero deep inside the superconductor. Accordingly, the function $\mathcal{D}^L_S(x,x',\varepsilon)$ becomes singular in this case.  Nevertheless, the combinations $\hat R_j^L(\mathcal{M}^L)^{-1}$ and $\hat R_{12}^L(\mathcal{M}^L)^{-1}$ remain finite also in this limit. At subgap energies we obtain
\begin{multline}
\hat R_1^L(\hat{\mathcal{M}}^L)^{-1}=
\hat R_2^L(\hat{\mathcal{M}}^L)^{-1}=
\hat R_{12}^L(\hat{\mathcal{M}}^L)^{-1}
\\=
\dfrac{1}{r_{N_1}+r_{N_2}+
\dfrac{2(\Delta^2-\varepsilon^2)e^{d/\xi_S(\varepsilon)}}{
\Delta^2 r_{\xi_S}(\varepsilon) G_{T_1} G_{T_2}}},
\end{multline}
where $d=|x_2-x_1|$ is the distance between two $SF$ contacts. 
Substituting the above relations into Eq.  \eqref{g12}
we arrive at the final result for the non-local spectral conductance of our device at subgap energies
\begin{widetext}
\begin{multline}
g_{12}(\varepsilon)=g_{21}(\varepsilon)=\dfrac{\Delta^2 - \varepsilon^2}{\Delta^2}
\dfrac{r_{\xi_S}(\varepsilon) \exp[-d/\xi_S (\varepsilon)]}{
2[r_{N_1} + 1/g_{T1}(\varepsilon)][r_{N_2} + 1/g_{T2}(\varepsilon)]}
\\
\times\left[
1+\bm{m}_1\bm{m}_2
\dfrac{G_{m1}}{g_{T1}(\varepsilon)}\dfrac{G_{m2}}{g_{T2}(\varepsilon)}
\dfrac{\Delta^2}{\Delta^2 - \varepsilon^2}
\dfrac{1}{1-\dfrac{\varepsilon^2}{\Delta^2}+ \dfrac{r_{N_1}+r_{N_2}}{2}
r_{\xi_S}(\varepsilon) G_{T_1} G_{T_2} \exp[-d/\xi_S(\varepsilon)] }
\right], \quad |\varepsilon|< \Delta.
\label{g12e}
\end{multline}
\end{widetext}

Eq. \eqref{g12e} represents the central result of our paper. It consists of two different contributions. The first of them is independent of the interface polarizations $\bm{m}_{1,2}$. This term represents direct generalization of the result \cite{Golubev09} in two different aspects.  Firstly, the analysis \cite{Golubev09}
was carried out under the assumption $r_{N_{1,2}}g_{T1,2}(\varepsilon) \ll 1$ which is abandoned here. Secondly (and more importantly), sufficiently large exchange fields $h_{1,2} \gg \Delta$ of ferromagnetic electrodes suppress disorder-induced electron interference in these electrodes and, hence, eliminate the corresponding zero-bias anomaly both in local \cite{VZK,HN,Z} and non-local \cite{Golubev09} spectral conductances. In this case with sufficient accuracy one can set $g_{Ti}(\varepsilon)=G_{Ti}\nu_S(x_i,\varepsilon)$ implying that at subgap energies
$g_{Ti}(\varepsilon)$ is entirely determined by the second term in Eq. (\ref{dos1})
which yields in the case of quasi-one-dimensional electrodes
\begin{gather}
g_{T1}(\varepsilon)=\dfrac{\Delta^2 G_{T_1}r_{\xi_S}(\varepsilon)}{2(\Delta^2-\varepsilon^2)}
\left[ G_{T_1} + G_{T_2}e^{-d/\xi_S (\varepsilon)} \right],
\\
g_{T2}(\varepsilon)=\dfrac{\Delta^2 G_{T_2}r_{\xi_S}(\varepsilon)}{2(\Delta^2-\varepsilon^2)}
\left[ G_{T_2} + G_{T_1}e^{-d/\xi_S (\varepsilon)} \right].
\end{gather}

Note, that
if the exchange field $h_{1,2}$ in both normal electrodes is reduced well below $\Delta$ and eventually is set equal to zero, the term containing $\hat U_1(\varepsilon )$ in Eqs. (\ref{g1T}), (\ref{g1L}) becomes important and should be taken into account. In this case we again recover the zero-bias anomaly \cite{VZK,HN,Z} $g_{Ti}(\varepsilon) \propto 1/\sqrt{\varepsilon}$ and from the first term in Eq. \eqref{g12e} we reproduce the results \cite{Golubev09} derived in the limit $h_{1,2} \to 0$.

\begin{figure}
\includegraphics[width=7.5cm]{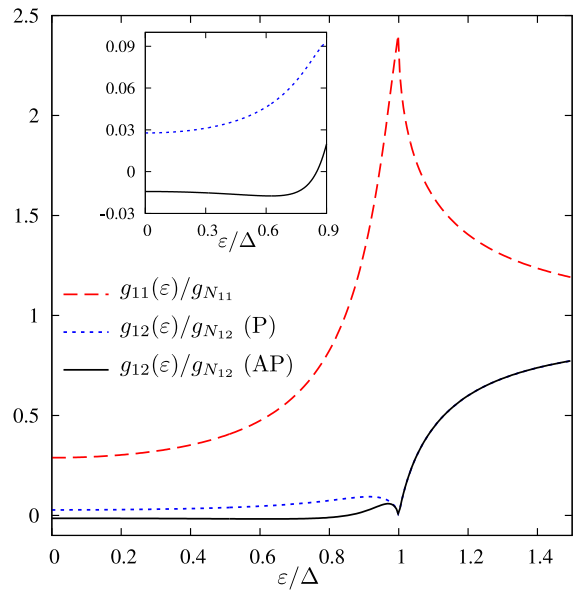}
\caption{(Color online) Local (long-dashed line) and non-local (short-dashed and solid lines) spectral conductances normalized to its normal state values. Here we choose $r_{N_1} = r_{N_2} = 5 r_{\xi_S}(0)$, $x_1 = L-x_2 = 5 \xi_S(0)$, $x_2-x_1=\xi_S (0)$, $G_{T_1} = G_{T_2} = 4G_{m_1} = 4G_{m_2}=0.2/r_{\xi_S}(0)$. Energy dependence of non-local conductance is displayed for parallel (P) $\bm{m}_1\bm{m}_2=1$ and antiparallel (AP) $\bm{m}_1\bm{m}_2=-1$ interface magnetizations. Inset: The same in the limit of low energies.}
\label{fig-e}
\end{figure}

The second term in Eq. \eqref{g12e} is proportional to the product $\bm{m}_1\bm{m}_2 G_{m1} G_{m2}$ and
describes non-local magnetoconductance effect in our system emerging due to spin-sensitive electron scattering at $SF$ interfaces. It is important that -- despite the strong inequality $|G_{mi}| \ll G_{Ti}$ -- both terms in Eq. \eqref{g12e} can be of the same order, i.e. the second (magnetic) contribution can significantly modify the non-local conductance of our device.

In the limit of large interface resistances $r_{N_{1,2}}g_{T1,2}(\varepsilon) \ll 1$ the formula \eqref{g12e} reduces to a much simpler one
\begin{multline}
g_{12}(\varepsilon)=g_{21}(\varepsilon)=
\dfrac{r_{\xi_S}(\varepsilon)}{2}
\exp[-d/\xi_S (\varepsilon)]
\label{g12e1}\\
\times\left[
\dfrac{\Delta^2 - \varepsilon^2}{\Delta^2}
g_{T1}(\varepsilon)g_{T2}(\varepsilon)+
\bm{m}_1\bm{m}_2 G_{m1} G_{m2}
\dfrac{\Delta^2}{\Delta^2-\varepsilon^2}\right].
\end{multline}
Interestingly, Eq. (\ref{g12e1}) remains applicable for arbitrary values of the angle between interface polarizations $\bm{m}_1$ and $\bm{m}_2$ and strongly resembles the analogous result for the non-local conductance in ballistic $FSF$ systems (cf., e.g., Eq. (77) in Ref. \onlinecite{Kalenkov07}). The first term in the square brackets in Eq. \eqref{g12e1} describes the fourth order contribution in the interface transmissions which remains nonzero also in the limit of the nonferromagnetic leads \cite{Golubev09}. In contrast, the second term is proportional to the product of transmissions of both interfaces, i.e. only to the second order in barrier transmissions \cite{FFH,Kalenkov07}. This term vanishes identically provided at least one of the interfaces is spin-isotropic.

Contrary to the non-local conductance at subgap energies, both local conductance (at all energies) and non-local spectral conductance at energies above the superconducting gap are only weakly affected by magnetic effects. Neglecting small corrections due to $G_m$ term in the boundary conditions we obtain
\begin{gather}
\hat g_{11}(\varepsilon)=\hat R_1^T(\hat{\mathcal{M}}^T)^{-1}, \quad
\hat g_{22}(\varepsilon)=\hat R_2^T(\hat{\mathcal{M}}^T)^{-1},
\label{g11c}
\\
\hat g_{12}(\varepsilon)=g_{21}(\varepsilon)=\hat R_{12}^T(\hat{\mathcal{M}}^T)^{-1}, \quad
|\varepsilon| > \Delta.
\label{g12c}
\end{gather}

Eqs. \eqref{g11c} and \eqref{g12c} together with the above expressions for the  non-local subgap conductance enable one to recover both local and non-local spectral conductances of our system at all energies. Typical energy dependencies for both $g_{11}(\varepsilon)$ and $g_{12}(\varepsilon)$ are displayed in Fig. \ref{fig-e}.
For instance, we observe that at subgap energies the non-local conductance $g_{12}$ changes its sign being positive for parallel and negative for antiparallel interface polarizations.

\begin{figure}
\includegraphics[width=7.5cm]{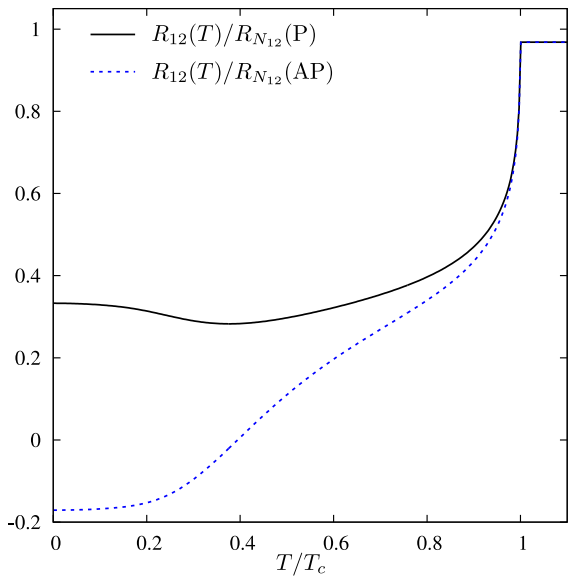}
\caption{(Color online) Non-local resistance (normalized to its normal state value) versus temperature (normalized to the superconducting critical temperature $T_C$) for parallel (P) and antiparallel (AP) interface magnetizations. The parameters are the same as in Fig. \ref{fig-e}.}
\label{fig-r}
\end{figure}

\begin{figure}
\includegraphics[width=7.5cm]{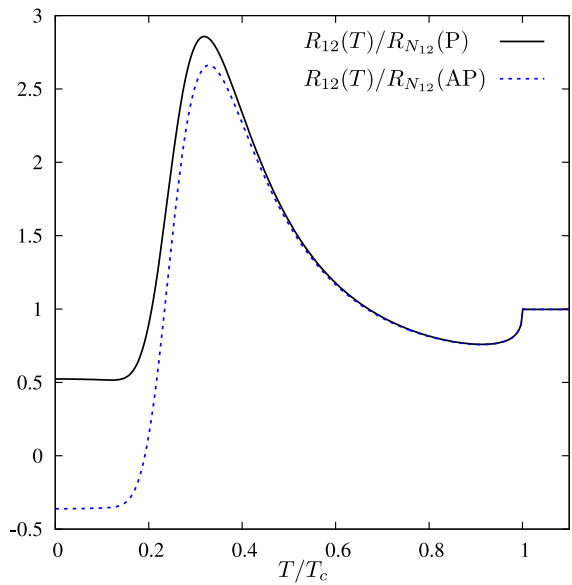}
\caption{(Color online) The same as in Fig. 3 for the following parameter values: $r_{N_1} = r_{N_2} = 5 r_{\xi_S}(0)$, $x_1 = L-x_2 = 5 \xi_S(0)$, $x_2-x_1=\xi_S (0)$, $G_{T_1} = G_{T_2} = 50G_{m_1} = 50G_{m_2}=0.025/r_{\xi_S}(0)$. }
\label{fig-r1}
\end{figure}

Having established the spectral conductance matrix $g_{ij}(\varepsilon)$ one can easily recover the complete $I-V$ curves for our hybrid $FSF$ structure. In the limit of low bias voltages these $I-V$ characteristics become linear, i.e.
\begin{gather}
I_1 = G_{11}(T) V_1 + G_{12}(T) V_2,
\label{I1}
\\
I_2 = G_{21}(T) V_1 + G_{22}(T) V_2,
\label{I2}
\end{gather}
where $G_{ij}(T)$ represent the linear conductance matrix defined as
\begin{equation}
G_{ij}(T)=\dfrac{1}{4T}
\int g_{ij}(\varepsilon)\dfrac{d\varepsilon}{\cosh^2\dfrac{\varepsilon}{2T}}.
\end{equation}
It may also be convenient to invert the relations \eqref{I1}-\eqref{I2} thus expressing induced voltages $V_{1,2}$ in terms of injected currents $I_{1,2}$:
\begin{gather}
V_1 = R_{11}(T) I_1 - R_{12}(T) I_2,
\label{V1}
\\
V_2 = -R_{21}(T) I_1 + R_{22}(T) I_2,
\label{V2}
\end{gather}
where the coefficients $R_{ij}(T)$ define local ($i=j$) and nonlocal ($i\neq j$) resistances
\begin{gather}
R_{11}(T)=\dfrac{G_{22}(T)}{G_{11}(T)G_{22}(T)-G_{12}^2(T)},
\\
R_{12}(T)=R_{21}(T)=\dfrac{G_{12}(T)}{G_{11}(T)G_{22}(T)-G_{12}^2(T)}
\end{gather}
and similarly for $R_{22}(T)$. In non-ferromagnetic $NSN$ structures the low temperature non-local resistance  $R_{12}(T\to 0)$ turns out to be independent of both the interface conductances and the parameters of the normal leads \cite{Golubev09}. However, this universality of $R_{12}$ does not hold anymore provided
non-magnetic normal metal leads are substituted by ferromagnets.  Non-local linear resistance $R_{12}$ of our $FSF$ structure is displayed in Figs. \ref{fig-r}, \ref{fig-r1} as a function of temperature for parallel ($\bm{m}_1\bm{m}_2=1$) and antiparallel ($\bm{m}_1\bm{m}_2=-1$) interface magnetizations. In Fig. \ref{fig-r} we show typical temperature behavior of the non-local resistance for sufficiently transparent interfaces. For both mutual interface magnetizations $R_{12}$ first decreases with temperature below $T_C$ similarly to the non-magnetic case. However, at lower $T$ important differences occur: While in the case of parallel magnetizations $R_{12}$ always remains positive and even shows a noticeable upturn at sufficiently low $T$, the non-local resistance for antiparallel magnetizations keeps monotonously decreasing with $T$ and may become negative in the low temperature limit. In the limit of very low interface transmissions the temperature dependence of the non-local resistance exhibits a well pronounced charge imbalance peak (see Fig. \ref{fig-r1}) which physics is similar to that analyzed in the case of non-ferromagnetic $NSN$ structures \cite{KZ06,GZ07,GKZ}. Let us point out that the above behavior of the non-local resistance is qualitatively consistent with available experimental observations \cite{Beckmann}.

\section{Concluding remarks}

In this paper we developed a quantitative theory of non-local electron transport in three-terminal hybrid ferromagnet-superconductor-ferromagnet structures in the presence of disorder in the electrodes. Within our model transfer of electrons across $SF$ interfaces is described in the tunneling limit and magnetic properties of the system
are accounted for by introducing ($i$) exchange fields $\bm{h}_{1,2}$ in both normal metal electrodes and ($ii$) magnetizations  $\bm{m}_{1,2}$ of both $SF$ interfaces (the model of spin-active interfaces). The two ingredients ($i$) and ($ii$) of our model
are in general independent from each other and have different physical implications. While the role of (comparatively large) exchange fields  $h_{1,2}\gg \Delta$ is merely to suppress disorder-induced interference of electrons \cite{VZK,HN,Z} penetrating from a superconductor into ferromagnetic electrodes, spin-sensitive electron scattering at $SF$ interfaces yields an extra contribution to the non-local conductance which essentially depends on relative orientations of the interface magnetizations. For anti-parallel magnetizations the total non-local conductance $g_{12}$ and resistance $R_{12}$ can turn negative at sufficiently low energies/temperatures. At higher temperatures the 
difference between the values of $R_{12}$ evaluated for
parallel and anti-parallel magnetizations becomes less important. At such temperatures the non-local resistance behaves similarly to the non-magnetic case demonstrating, e.g., a well-pronounced charge imbalance peak \cite{GKZ} in the limit of low interface transmissions.

We believe that our predictions can be directly used for quantitative analysis of future experiments on non-local electron transport in hybrid $FSF$ structures.

\vspace{0.5cm}

\centerline{\bf Acknowledgments}

\vspace{0.5cm}

This work was supported in part by DFG and by RFBR grant 09-02-00886. M.S.K. also acknowledges support from the Council for grants of the Russian President (Grant No. 89.2009.2) and from the Dynasty Foundation.


\begin{thebibliography}{99}
\bibitem{And} A.F. Andreev, Sov. Phys. JETP {\bf 19}, 1228 (1964).
\bibitem{car} J.M. Byers and M.E. Flatte, Phys. Rev. Lett. {\bf 74}, 306 (1995);
G. Deutscher and D. Feinberg, Appl. Phys. Lett. {\bf 76}, 487
(2000).
\bibitem{FFH} G. Falci, D. Feinberg, and F.W.J. Hekking, Europhys. Lett. \textbf{54}, 255 (2001).
\bibitem{KZ06} M.S. Kalenkov and A.D. Zaikin, Phys. Rev. B \textbf{75}, 172503 (2007); JETP Lett. {\bf 87}, 140 (2008).
\bibitem{Beckmann} D. Beckmann, H.B. Weber, and H. v. L\"ohneysen, Phys. Rev. Lett. \textbf{93}, 197003
(2004).
\bibitem{Teun} S. Russo, M. Kroug, T.M. Klapwijk, and A.F. Morpurgo, Phys. Rev. Lett. \textbf{95}, 027002 (2005).
\bibitem{Venkat} P. Cadden-Zimansky and V. Chandrasekhar, Phys. Rev. Lett. \textbf{97}, 237003
(2006); P. Cadden-Zimansky, Z. Jiang, and V. Chandrasekhar, New J.
Phys. {\bf 9}, 116 (2007).
\bibitem{Basel} A. Kleine, A. Baumgartner, J. Trbovic, and C. Sch\"onenberger, Europhys. Lett. {\bf 87}, 27011 (2009); A. Kleine, A. Baumgartner, J. Trbovic, D.S. Golubev, A.D. Zaikin, and C. Sch\"onenberger, arXiv:0911.4427 (2009).
\bibitem{Deutscher} B. Almog, S. Hacohen-Gourgy, A. Tsukernik, and
G. Deutscher, Phys. Rev. B \textbf{80}, 220512(R) (2009).
\bibitem{Beckmann2} J. Brauer, F. H\"ubler, M. Smetanin, D. Beckmann, and H. v. L\"ohneysen, Phys. Rev. B {\bf 81}, 024515 (2010).
\bibitem{LY} A. Levy Yeyati, F.S. Bergeret, A. Martin-Rodero, and T.M. Klapwijk,  Nat. Phys. {\bf 3}, 455 (2007).
\bibitem{GZ09} D.S. Golubev and A.D. Zaikin, Europhys. Lett. \textbf{86}, 37009 (2009).
\bibitem{BG} A. Brinkman and A.A. Golubov, Phys. Rev. B
\textbf{74}, 214512 (2006).
 \bibitem{Belzig} J.P. Morten, A. Brataas, and W. Belzig, Phys. Rev. B \textbf{74}, 214510 (2006).
\bibitem{Melin} R. Melin, Phys. Rev. B \textbf{73}, 174512 (2006).
\bibitem{GZ07} D.S. Golubev and A.D. Zaikin, Phys. Rev. B \textbf{76}, 184510 (2007).
\bibitem{Golubev09} D.S. Golubev, M.S. Kalenkov, and A.D. Zaikin, Phys. Rev. Lett. {\bf 103}, 067006 (2009).
\bibitem{Kalenkov07} M.S. Kalenkov and A.D. Zaikin, Phys. Rev. B {\bf 76}, 224506
(2007).
\bibitem{Kalenkov07E} M.S. Kalenkov and A.D. Zaikin, Physica E {\bf 40}, 147 (2007).
\bibitem{BWBSZ} See, e.g.,
W. Belzig, F. Wilhelm, C. Bruder, G. Sch\"on, and A.D. Zaikin,
Superlatt. Microstruct. \textbf{25}, 1251 (1999).
\bibitem{Eschrig} M. Eschrig, Phys. Rev. B {\bf 80}, 134511 (2009).
\bibitem{Eschrig2} M. Eschrig, J. Kopu, J. C. Cuevas, and G. Sch\"on, Phys. Rev.
Lett. {\bf 90}, 137003 (2003); M. Eschrig and T. Lofwander, Nat.
Phys. {\bf 4}, 138 (2008).
\bibitem{GKZ} A.V. Galaktionov, M.S. Kalenkov, and  A.D. Zaikin, Phys. Rev. B
{\bf 77}, 094520 (2008); M.S. Kalenkov, A.V. Galaktionov, and A.D.
Zaikin, Phys. Rev. B {\bf 79}, 014521 (2009).
\bibitem{Huertas02} D. Huertas-Hernando, Yu.V. Nazarov, and W. Belzig, Phys. Rev. Lett. \textbf{88}, 047003 (2002);
A. Cottet, D. Huertas-Hernando, W. Belzig, and Yu.V. Nazarov,
Phys. Rev. B \textbf{80}, 184511 (2009).
\bibitem{VZK} A.F. Volkov, A.V. Zaitsev, and T.M. Klapwijk, Physica C \textbf{210}, 21 (1993).
\bibitem{HN} F.W.J. Hekking and Yu.V. Nazarov, Phys. Rev. Lett. \textbf{71}, 1625 (1993).
\bibitem{Z} A.D. Zaikin, Physica B \textbf{203}, 255 (1994).
\end{thebibliography}
\end{document}